\documentclass[prb,aps,reprint,showpacs,amssymb, superscriptaddress,footinbib]{revtex4-1}
\usepackage{graphicx}
\usepackage{amsmath}
\usepackage[utf8]{inputenc}
\usepackage{amsmath}
\usepackage{amsfonts}
\usepackage{amssymb}
\usepackage{graphicx}
\usepackage{forloop}
\usepackage[multidot]{grffile}
\usepackage{color}   
\usepackage{float}
\usepackage{multirow}
\usepackage[section]{placeins}

\begin{document}
\title{Interplay between exotic superfluidity and magnetism in a chain of four-component ultracold atoms}

\author{E. Szirmai}
\email[electronic address: ]{eszirmai@gmail.com}
\affiliation{BME-MTA Exotic Quantum Phases "Lend\"ulet" Research Group, Budapest University of Technology and Economics, Institute of Physics, H-1111 Budapest, Hungary}

\author{G. Barcza}
\email[electronic address: ]{barcza.gergely@wigner.mta.hu}
\affiliation{Strongly Correlated Systems "Lend\"ulet" Research Group, Wigner Research Centre for Physics, HAS,  H-1525 Budapest, Hungary}

\author{J. S\'olyom}
\affiliation{Strongly Correlated Systems "Lend\"ulet" Research Group, Wigner Research Centre for Physics, HAS,  H-1525 Budapest, Hungary}

 \author{\"O. Legeza}
\affiliation{Strongly Correlated Systems "Lend\"ulet" Research Group, Wigner Research Centre for Physics, HAS,  H-1525 Budapest, Hungary}

\date\today

\begin{abstract}
We investigate  the spin-polarized chain of ultracold fermionic atoms with  spin-3/2 described by the fermionic Hubbard model with SU(4) symmetric attractive interaction. The competition of  bound pairs, trions, quartets and unbound atoms is studied analytically and by density matrix renormalization group simulations. 
We find several distinct states where bound particles coexist with the ferromagnetic state of unpaired fermions.
In particular, an exotic inhomogeneous Fulde-Ferrell-Larkin-Ovchinnikov (FFLO)-type superfluid of quartets in a magnetic background of uncorrelated atoms is found for weaker interactions.
We show that the system can be driven from this quartet-FFLO state to a molecular state of localized quartets which is also reflected in the static structure factor.  
For strong enough coupling, spatial segregation between molecular crystals and ferromagnetic liquids emerges due to the large effective mass of the composite particles. 
\end{abstract}

\pacs{71.30.+h, 71.10.Fd}

\maketitle

\textit{Introduction.} --- Investigating effective Hamiltonians with local contact interactions has proven to be instrumental in understanding the physics of ultracold atomic systems.\cite{Lewenstein2007,Bloch2008,Ketterle2008a}
Realizing exotic quantum states, e.g., inhomogeneous FFLO superfluid  pairs \cite{Fulde1964,Larkin1964} or trionic state \cite{Rapp2007}, low-dimensional systems are of particular interest owing to the large quantum fluctuations.\cite{Yang2001,Recati2003,Lecheminant2005,Guan2009,Liao2010,Manmana2011,Duivenvoorden2013} 
Thanks to the rapid progress in the experimental techniques in the physics of ultracold atoms, by now not only two-component  systems can be  experimentally realized but an insight into unconventional molecular superfluids of multiple-body states is also provided.
\cite{Zwierlein2006,Partridge2006,Taie2010,Desalvo2010,Liao2010,Gorshkov2010,Krauser2012,Nascimbene2012,Scazza2014,Cappellini2014,Zhang2014,Greif2014} 
Among others, four-component systems could be experimentally investigated due to the four hyperfine states of high-spin alkali and alkaline-earth atoms, such as $^6$Li, $^{40}$K,  $^{123}$Cs and $^9$Be, respectively.\cite{Wu2006,Lewenstein2012,Cazalilla2014} 
Additionally, ground-breaking recent experiments on fermionic atoms  with tunable spin, e.g., on $^{87}$Sr atoms \cite{Stellmer2011} and  on $^{173}$Yb  loaded in one-dimensional array,\cite{Pagano2014a} urge further studies of the detailed properties of such systems.

The spin-neutral phase diagram of a one-dimensional four-component interacting Fermi gas with $s$-wave scattering  exhibiting various exotic superfluid phases like  SU(4)-singlet quartets is well established.\cite{Wu2005, Wu2006,Capponi2007,Lecheminant2008,Roux2009,Capponi2015} At the same time, for finite magnetic polarization
\begin{equation}
\label{eq:polarization}
p=\frac{1}{L}\sum_{i,\alpha} \alpha \langle \hat{n}_{\alpha,i} \rangle\,,
\end{equation}
our knowledge is rather poor. [In Eq.~\eqref{eq:polarization} we use the standard notation of the particle number operator $\hat{n}_{\alpha,i}$ at site $i$ carrying spin $\alpha\in \{ -3/2,-1/2,1/2,3/2 \}$, and $L$ denotes the length of the chain.] In presence of spin imbalance, it is expected that the singlet quartets are replaced at least partially by other energetically favorable spin-carrying excitations. 
 In our earlier works\cite{Barcza2012,Barcza2015} we showed that if the attractive interaction is far detuned from SU(4) symmetry, a mixture of spin-carrying pairs and spin-neutral quartets develops.
The weight of the pairs increases  monotonically as a function of the polarization at the expense of progressively eliminating quartets which disappear completely for $p=1$.
Moreover, we showed that the quasi-condensates can spatially segregate 
providing a clear domain structure of the exotic superfluid mixture consisting of spin-carrying BCS-like pairs with maximum spin polarization and SU(4) singlet quartets. 

A different behavior is expected for the experimentally intriguing SU(4) symmetric interaction which can be achieved by $s$-wave scatterings between alkaline-earth atoms. To a good accuracy the scattering does not depend on the $F=3/2$ hyperfine spin  of the colliding atoms, consequently the interaction shows an SU(2$F$+1)=SU(4) symmetry. \cite{Gorshkov2010} 
In Ref.~\onlinecite{Guan2009}, investigating the SU(4) delta-gas by the Bethe Ansatz, a smooth transition  was reported from a quartetting state to a normal Fermi liquid state of unbound fermions as the quartets break up one-by-one for increasing spin imbalance.

In this paper, we study the competition between magnetism and superfluidity by controlling the spin polarization of the four-component quarter-filled Hubbard model through the linear Zeeman term, $pNB$, where $N$ is the total number of particles, $B$ denotes the external magnetic field, and the magnetic polarization is given in Eq.~\eqref{eq:polarization}.
\begin{equation}
{\cal H} = -t\sum_{i,\alpha}^L \big( \hat{c}_{\alpha,i}^\dagger \hat{c}_{\alpha,i+1}^{\phantom\dagger} + {\rm H.c. }\big)   + \frac{U}{2} \sum_{i=1}^L \hat{n}_i^2 + pNB 
 \label{eq:ham1}
\end{equation}
Here, $\hat{c}_{\alpha,i}^\dagger$  creates a fermion with spin $\alpha$ at site $i$ and $\hat{n}_i=\sum_\alpha \hat{n}_{\alpha,i}$ is the number operator, $t$ measures the one-particle overlap between neighboring sites and $U<0$ parametrizes the strength of the attraction.

\begin{figure}[tb]
\includegraphics[scale=0.72]{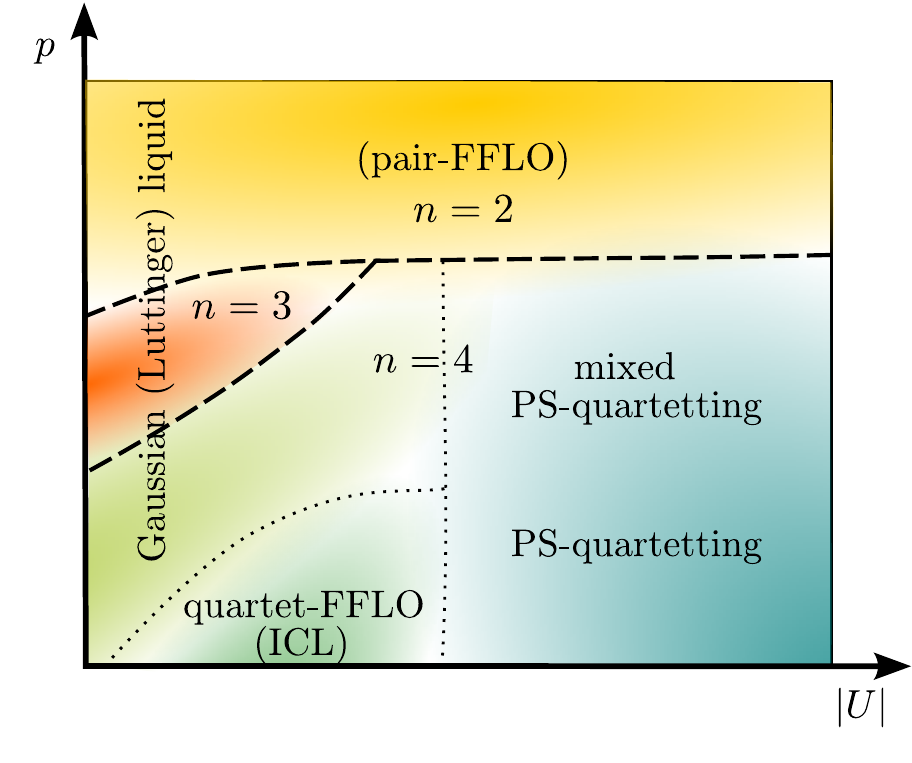}
\caption{(Color online) Schematic phase diagram of spin-3/2 fermionic atoms in case of spin-independent interaction as a function of the strength of the attractive interaction $|U|$ and the polarization $0<p<3/2$. 
The dashed lines indicate transitions between phases consisting different number of spin components $n=2$, 3, and 4. 
The $n=1$ fully polarized phase is not indicated.
The lines are only guide to the eyes.
}
\label{fig:phasediagram}
\end{figure}

The quarter filled, $N=L$,  phase diagram based on our  density matrix renormalization group (DMRG) \cite{White1992,White1993} calculations and bosonization study is summarized in Fig.~\ref{fig:phasediagram} as a function of the coupling $U$ and polarization $p$.
The partially polarized system exhibits three quartetting phases  with strikingly distinct character for increasing $|U|$ in the four-component region. 
Above a critical polarization only $\alpha >0$ fermions are present in the system, i.e., we observe a gapless two-component liquid.
We found that three-component system emerges for weak or intermediate interactions, while a direct transition between phases with  four and two fermion components is observed in case of strong enough couplings.
In what follows we first give a qualitative description of the behavior of the model in the bosonization representation and then we present the results of numerical calculations using the DMRG method.

\textit{Bosonization.} --- The properties determined by low-energy excitations can be well understood within hydrodynamical treatment taking into account only those states which are in the vicinity of the Fermi energy. In case of spin balance the population of the 4 spin states are equal, and there are only two Fermi points $\pm k_\mathrm{F}$.

Switching on spin imbalance this degeneracy is removed and the Fermi momenta shift leading to eight different Fermi points at $\pm k_\alpha$'s. 
Using standard notations,\cite{Gogolin2004,Giamarchi2004} in boson representation the total Hamiltonian can be written in the form
\begin{multline}
\label{eq:ham-bos}
H=  \frac{1}{\pi} \sum_\alpha \int \mathrm{d} x  \,\,\bigg[  \left( 2v_\alpha + \frac{U}{\pi}  \right)   (\partial_x \phi_{\alpha})^2 +  2v_\alpha   (\partial_x \theta_{\alpha})^2  \\ 
+ \frac{U}{\pi} \sum_{\alpha'(\neq \alpha)} \partial_x \phi_{\alpha} \partial_x 
\phi_{\alpha'} 
+ \lambda_\alpha  \partial_x \phi_{\alpha} \\ 
+ \frac{U}{2\pi} \sum_{\alpha'} \mathrm{cos}\left[ 2(\phi_\alpha - \phi_{\alpha'})  \right]  + g_u  \sum_{\alpha'} \mathrm{cos}\Big[ 2 \sum_\alpha \phi_\alpha  \Big] \bigg] ,
\end{multline}
where the Gaussian part describes the small-momentum transfer scattering processes, while the cosine terms describe the ones with large-momentum transfer. $g_u$ drives the four-particle umklapp processes which may appear at quarter filling and $g_u$ is proportional to $U^3$. We introduce a spin-dependent Lagrange multiplier $\lambda_\alpha$ to fix the global number of the spin components and the total particle number per site. The two constraints can be ensured with two independent Lagrange multipliers $\lambda_n$, and $\lambda_s$ with $\lambda_\alpha = \lambda_n + \alpha \lambda_s$. With this the explicit expression of the shifted Fermi momenta are: $k_\alpha = k_\mathrm{F} + \mu_\mathrm{B} \alpha \lambda_s $, where $\mu_\mathrm{B}$ is the Bohr magneton, and $\alpha=\pm 1/2, \pm 3/2$. After the diagonalization of the Gaussian part of the Hamintonian~\eqref{eq:ham-bos} in the spin space, the linear term $\lambda_\alpha \partial_x\phi_\alpha$ can be transformed out by a simple shift of the new bosonic fields. In the following we discuss the properties of both the gapless, and gapped phases. We assume semi-classical behavior in presence of a gap, and characterize the various phases by the asymptotic behavior of the competing correlations.\cite{Gogolin2004,Giamarchi2004}

When the interaction is weak compared to the energy contribution of the magnetization and only particles close to the Fermi points can be excited and can participate in scattering processes, the quasi-momentum conservation cannot be satisfied by these relevant particles in  the large-momentum transfer scattering processes.\cite{Gogolin2004} 
This means that the spin imbalance freezes out the real backscatterings and one arrives at a Gaussian problem.  
In case of quarter filling, similarly to  the backward scatterings, the four-particle umklapp processes are also frozen out for weak interactions. 
In this case all the bosonic fields $\phi$ and their duals $\theta$ can fluctuate freely, and the dominant instability is a $2k_\mathrm{F}$-density order ($\mathcal{O}_{2k_\mathrm{F}} \propto e^{i2\phi_\alpha}$) with subdominant pairing, and quartetting. 
Further increase of the spin-polarization  freezes out the spin components one-by-one, and the system becomes equivalent to a three- and a two-component system, respectively. 
In principle, these transitions are not necessarily simple, i.e., close to the transition points a gap can be expected to pin the density fluctuations, whenever the backward scatterings become sufficiently strong as we discussed above.
However, we will see that our numerical results predict  that the system stays in the Gaussian state even close to the transition lines. 
In Fig.~\ref{fig:phasediagram} this phase is referred to as Gaussian (Luttinger) liquid with different color for the different number of the spin component $n$.
Further increasing the polarization, it obviously saturates at $p=3/2$, and a liquid state of unbound fermions with maximum spin projection ($\alpha=3/2$) is stabilized. This trivial, $n=1$, fully polarized phase is not indicated in Fig.~\ref{fig:phasediagram}.

\begin{figure}[htb]
\includegraphics[scale=0.23]{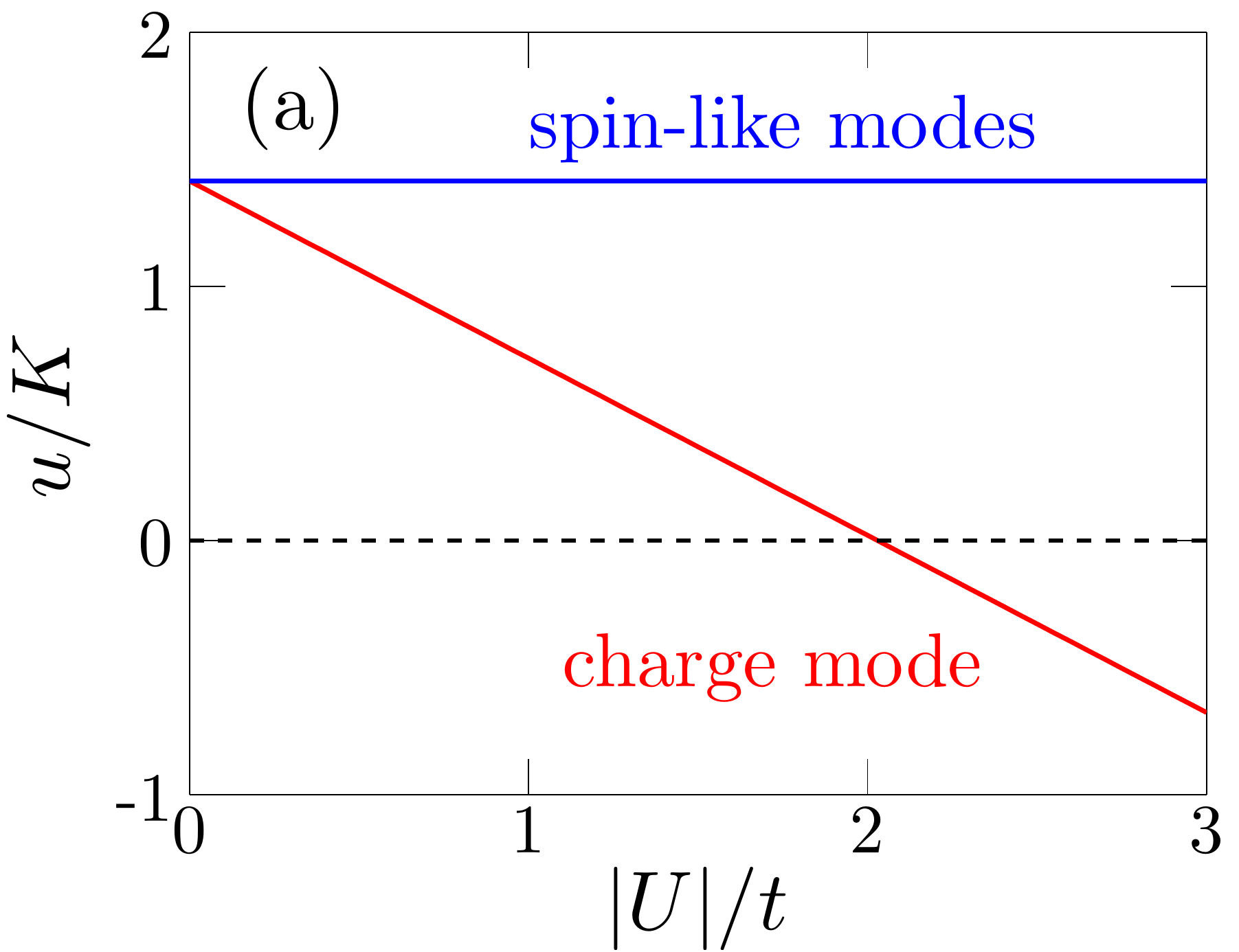}
\includegraphics[scale=0.23]{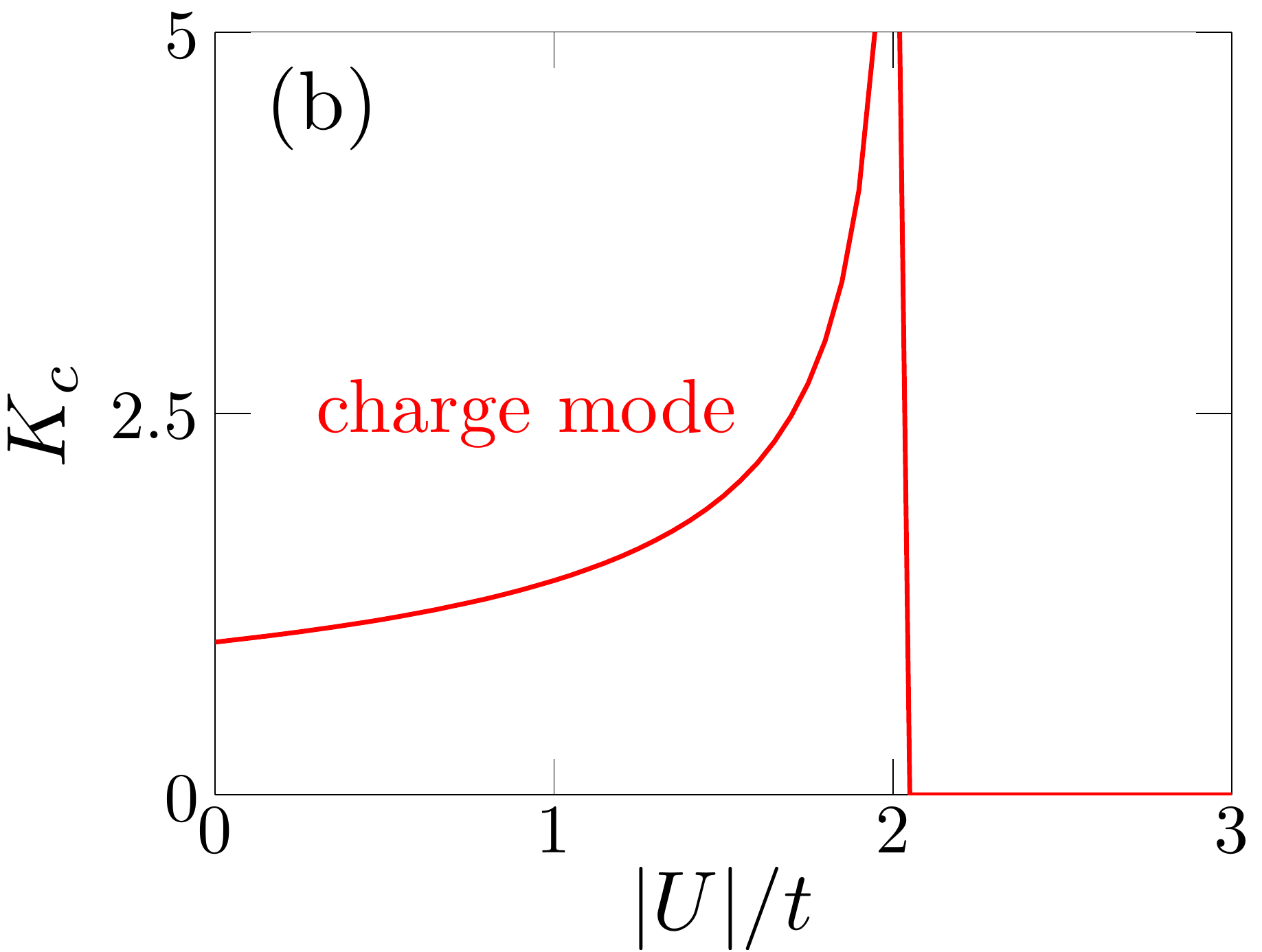}
\caption{(Color online) a) Eigenvalues of the Gaussian part of the Hamiltonian ~\eqref{eq:ham-bos} as a function of the coupling $U$, and b) the divergence of the Luttinger parameter of the charge mode $K_c$ indicating the density instability which leads to phase separation in the system.}
\label{fig:luttparams}
\end{figure}

An analysis of  the Luttinger parameters indicates that for sufficiently strong spin polarizations the increasing attractive interaction drives the system to a phase separated state via a first order transition. In spin-balanced system such a transition is characterized by the divergence of the compressibility $\kappa$ which is proportional to the Luttinger parameter $K_c$ of the charge mode: $\kappa \sim K_c/u_c$, where $u_c$ is the corresponding sound velocity. In presence of spin imbalance, the charge-like mode does not correspond to the symmetric combination of the density of the spin components $n_\mathrm{tot} = \sum_\alpha n_\alpha$, but a weighted sum of them: $n_\mathrm{tot}^\mathrm{w} = \sum_\alpha a_\alpha n_\alpha$ with $a_\alpha = \sqrt{\frac{\sin(k_\alpha)}{\sin(\pi/4)}}$. However, the compressibility is directly related to $n_\mathrm{tot}$, a singularity in the Luttinger parameter of the charge-like weighted mode also indicates instability in the system. Introducing the new fields $\phi_\alpha \rightarrow a_\alpha \phi_\alpha$ the three-fold degenerate eigenvalue of the Gaussian part of the Hamiltonian~\eqref{eq:ham-bos} corresponds to the spin-like modes, while the non-degenerate eigenvalue gives the ratio of the sound velocity and the Luttinger parameter of the charge-like mode $u_c/K_c$. The critical interaction strength is determined by the equation $u_c/K_c=0$. Fig.~\ref{fig:luttparams} shows the characteristic behavior of the four eigenvalues, and the Luttinger parameter $K_c$. We find that $K_c$ becomes divergent at a critical interaction strength. Although, the critical coupling is beyond the validity of the bosonic representation of the fermion model, the divergence of $K_c$ indicates density instability, and the possibility of phase separation for stronger interactions. This transition is indicated in Fig.~\ref{fig:phasediagram} by a dotted line between the Gaussian liquid of the 4-component system and the phase-separated (PS) quartetting phase.

In case of intermediate couplings one has to deal with the backscatterings, and the four-particle umklapp processes, too. Due to the different occupations of the four spin components, all the four bosonic fields are coupled by the cosine terms in Eq.~\eqref{eq:ham-bos}. Accordingly, a relevant backscattering or umklapp process can affect all the four modes opening a gap in their excitation spectra and leading to an incompressible liquid (ICL) state. 
In this case only the dual fields $\theta$ fluctuate freely, therefore the competing correlations are the tunneling density $\mathcal{O}_G \propto e^{i2 \theta_\alpha}$, and the various composite states: pairing $\mathcal{O}_P \propto e^{i2(\theta_\alpha+\theta_{\alpha'})}$, trioning $\mathcal{O}_T \propto e^{i2(\theta_{\alpha_1}+\theta_{\alpha_2}+\theta_{\alpha_3})}$, and quartetting $\mathcal{O}_Q \propto e^{i2\sum_\alpha \theta_\alpha}$. However, the tunneling density shows always the slowest decay, the system can be characterized by the off-diagonal orders of the competing composite bound states. 
For weak spin imbalance, the interaction is sufficiently strong to suppress the magnetic energy. In this regime our numerical analysis supports that the system is indeed a gapped incompressible liquid with dominant quartet correlations. The corresponding phase is indicated in Fig.~\ref{fig:phasediagram} as quartet-FFLO (ICL) which notation is based on the details of our numerical findings discussed below.

\textit{Numerical results.} --- In order to establish a complete phase diagram of the system we performed numerical calculations 
using DMRG method with open boundary condition up to $L=72$ sites.
The accuracy was controlled by the dynamic block state selection procedure (DBSS) \cite{Legeza2003,Legeza2004} keeping up to 2000 block states and performing 8 sweeps.
In addition to the various correlation functions and their Fourier transforms, we study local densities measuring the exclusive occupation number of unbound atoms and various molecules composed of two, three, or four fermions.
As examples, the explicit formula for $\alpha=3/2$ free atom,  quintet pair with $m=2$, trion with  $\alpha=3/2$ and quartet read
\begin{eqnarray}
A_{3/2,i}&=&\left\langle (1-\hat{n}_{-3/2,i})(1-\hat{n}_{-1/2,i})(1-\hat{n}_{1/2,i})\hat{n}_{3/2,i}\right\rangle ,\nonumber\\
P_{2,2,i}&=&\left\langle (1-\hat{n}_{-3/2,i})(1-\hat{n}_{-1/2,i})\hat{n}_{1/2,i}\hat{n}_{3/2,i}\right\rangle ,\nonumber\\
T_{3/2,i}&=&\left\langle (1-\hat{n}_{-3/2,i})\hat{n}_{-1/2,i}\hat{n}_{1/2,i}\hat{n}_{3/2,i}\right\rangle ,\nonumber\\
Q_i&=&\left\langle \hat{n}_{-3/2,i}\hat{n}_{-1/2,i}\hat{n}_{1/2,i}\hat{n}_{3/2,i}\right\rangle ,
\end{eqnarray}
respectively. Results presented in the  following are based on detailed calculations for: $U/t \in \{-0.1$, $-0.5$ ,$-1$, $-2$, $-3$, $-4$, $-8$, $-20$, $-50\}$.

Starting from the unpolarized case ($p=0$) our numerical study confirms the earlier results\cite{Wu2005,Wu2006, Roux2009,Capponi2007} about the gapless Gaussian state in the very weak interaction regime, and the formation of quartets for moderate attractive interactions. However, for extreme strong interactions --- far beyond the validity of the field-theoretical description --- we found that the quartets preferably behave classically and become localized as the effective repulsion between neighboring quartets, $\sim t^2/|U|$, overwhelms their effective hopping amplitude, $\sim t^4/|U|^3$. 

Polarizing the weakly interacting system up to moderate values, the ground state does not change drastically compared to the spin-balanced case. 
The correlation functions still show power-law decay, and the system remains in the Gaussian state as the phase diagram  in Fig.~\ref{fig:phasediagram} shows. 
However, as the polarization is increased, the weight of the $m=2$ pairs formed of $\alpha=3/2$, and $1/2$ fermions  increases and their correlations  become dominant over quartet decay. 
For sufficiently large polarizations, where the gapless system contains exclusively $\alpha>0$ particles and is therefore half-filled, the Green's function of the majority component ($\alpha=3/2$) is  dominant.
Nevertheless the  slow algebraic decay of $m=2$ pair correlations  with spatial oscillation periodicity proportional to the population imbalance  can be interpreted as FFLO pairing \cite{Feiguin2007} (also depicted in Fig.~\ref{fig:phasediagram}).

\begin{figure}[!tb]
   \includegraphics[scale=0.24]{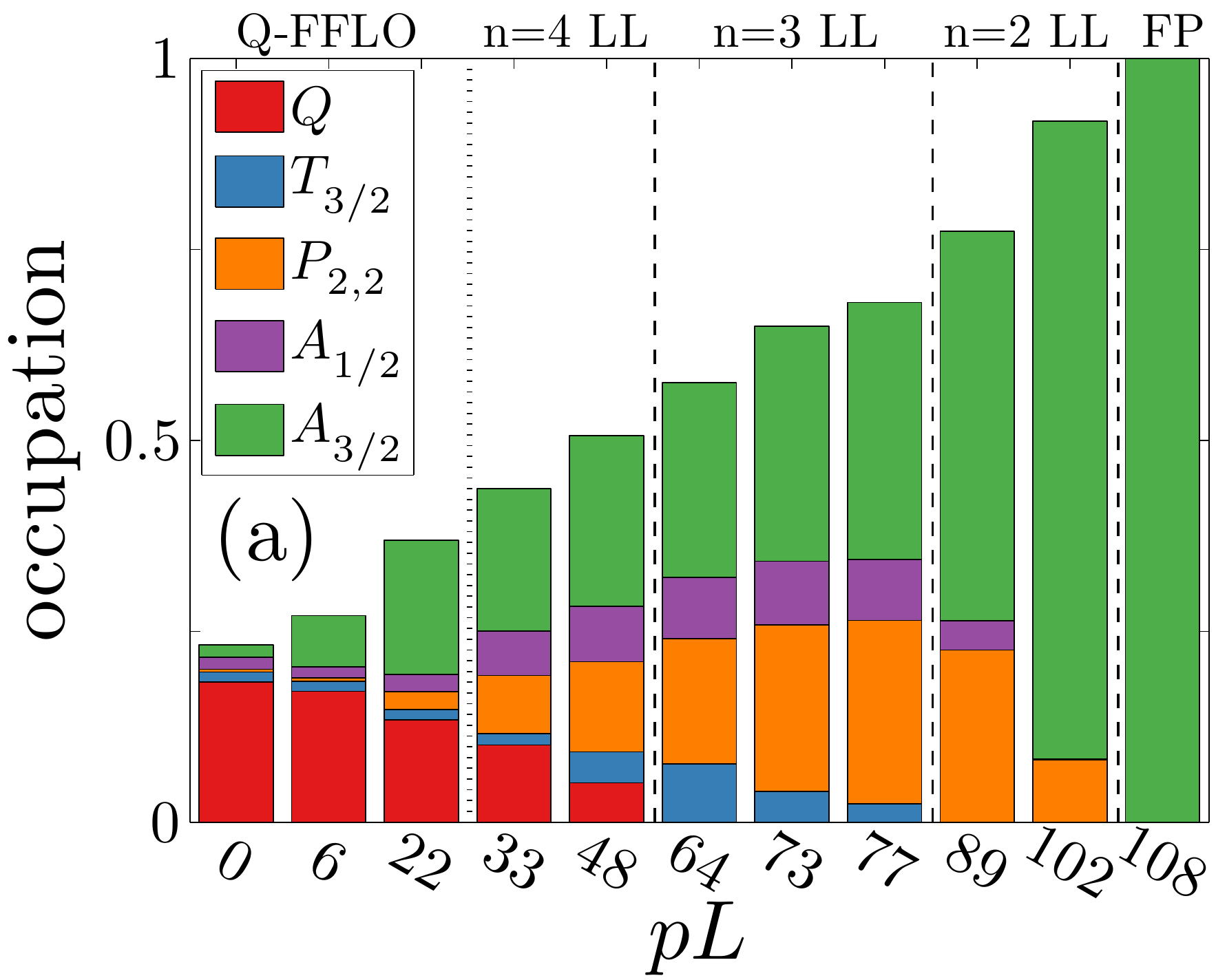} 
   \includegraphics[scale=0.24]{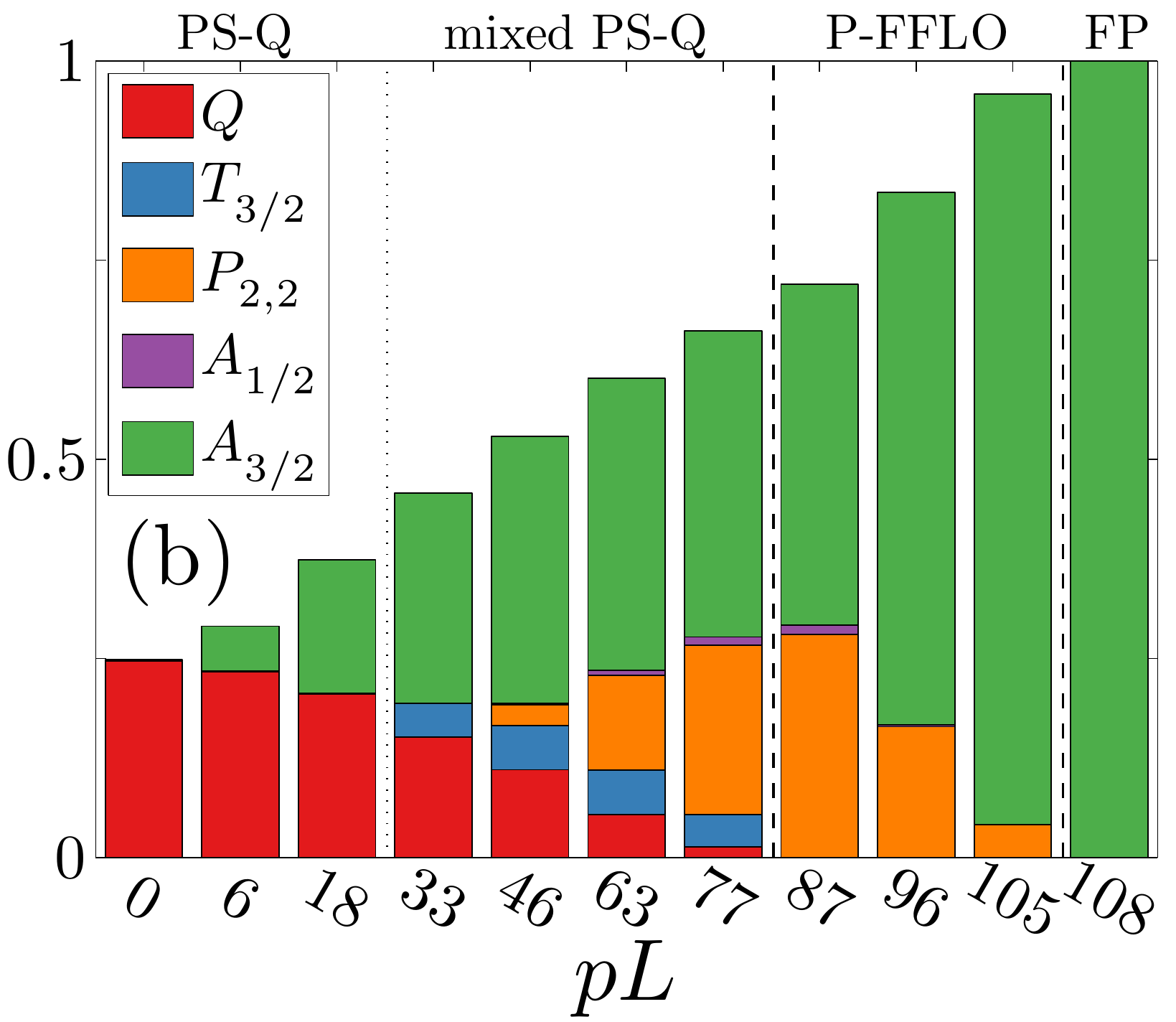}
\caption{(Color online)  a) Average occupation number of the characteristic composite particles in the chain with $L=72$ sites measured in representative $p$ values for $U/t=-2$. 
b)  Similar for  $U/t=-8$. 
}
\label{fig:densityintegral_U-2}
\end{figure}

For intermediate attractive interactions, where quartetting becomes more prominent, the SU(4)-singlet quartets of the unpolarized ground state start to dissolve into a mixture with $\alpha=3/2$ unpaired fermions as the spin balance is broken (see Fig. \ref{fig:densityintegral_U-2}). 
The nature of the quartets can be caught by the analysis of their center of mass (COM) momentum distribution.
An exotic FFLO-like state is found since the peak of the distribution shifts linearly with increasing polarization.\cite{Guan2013} 
This quartet-FFLO state can be identified to the incompressible liquid state of the bosonization predictions as it is shown in Fig.~\ref{fig:phasediagram}.
Note that in this state trions and $m=2$ quintet pairs can also be observed with small weight as shown in Fig.~\ref{fig:densityintegral_U-2} (a) for smaller values of $p$. 
As the polarization is further increased all the correlation functions start to show algebraic decay indicating that the gap is closing and the system becomes four-component Gaussian (see Fig.~\ref{fig:phasediagram}). Now again we found that the polarization does not affect the correlation functions in the Gaussian state, and up to intermediate couplings the spin components are frozen out one-by-one. 

Turning now to the strongly interacting regime, we observe that  in the polarized system  the quartet crystallization appears for significantly weaker $U$ compared to case $p=0$.
This can be understood as follows.
The spin-balanced ground state is  purely characterized by quartets and their localization is owed to the strong effective repulsion as discussed above.
Contrary to this, in case of spin imbalance, the crystallization and the spatial segregation of the emerging composite particles is governed by their mass imbalance similarly to  two-component systems with asymmetric hoppings.\cite{Barbiero2010,Guglielmino2011,Roscilde2012} 
In the present model, the asymmetry can be understood in terms of the perturbation theory: the effective hopping for pairs, trions, quartets is of order $t^2/|U|$, $t^3/|U|^2$, $t^4/|U|^3$, respectively.\cite{Roux2009}

\begin{figure}[!tb]
    \includegraphics[scale=0.245]{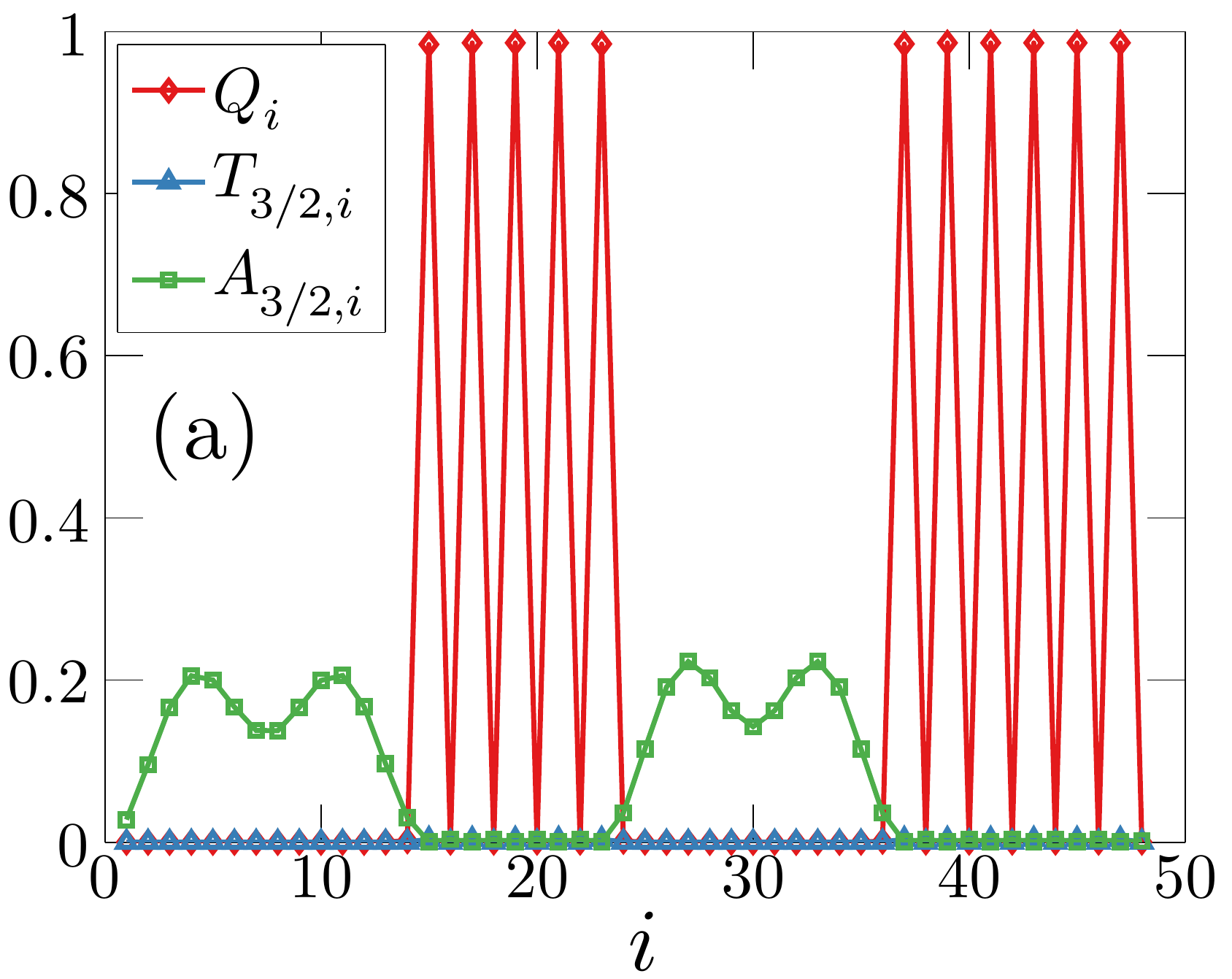}
    \includegraphics[scale=0.245]{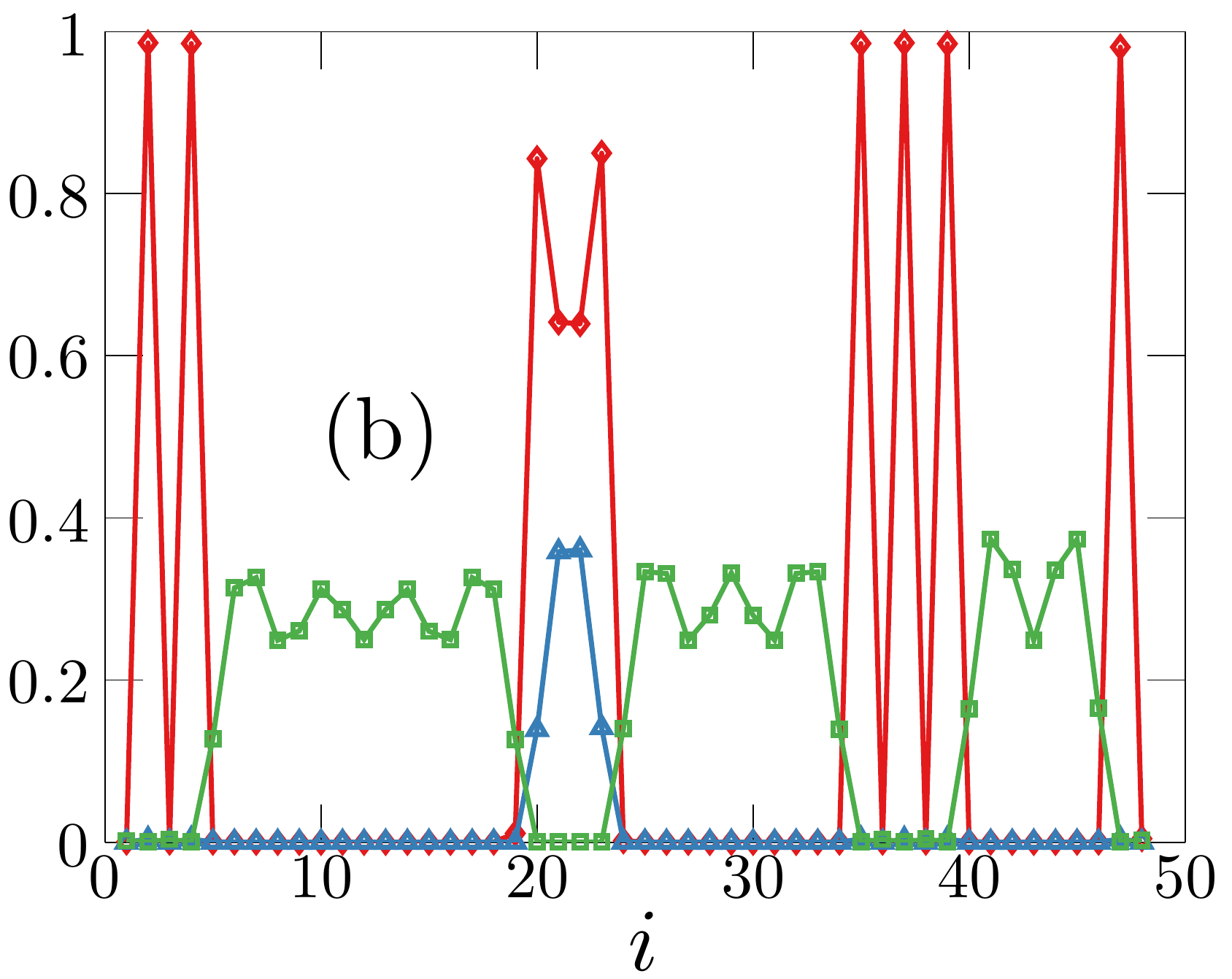}
\caption{(Color online) a) Typical density profile of the characteristic composite particles measured in the PS-quartetting phase  ($U/t=-8$, $p=1/8$).
b) Similar in the mixed PS-quartetting phase ($U/t=-8$, $p=5/16$).
Here, for better visibility, we present the results for a chain of length $L=48$. The lines are only guide to the eyes.}
\label{fig:density_U-8}
\end{figure}

In particular, for weak spin imbalance (PS-quartetting in Fig.~\ref{fig:phasediagram})  quartets become crystallized and segregate from the sea of unbound $\alpha=3/2$ atoms according to the density profiles (see Fig.~\ref{fig:density_U-8} (a)).
 The COM of the quartets remains zero independently of $p$ and they behave like well localized hard-core bosons.
For larger polarization, besides the free atoms, $\alpha=3/2$ trions  appear gradually as well.
The heavy particles  form  domains  so that the mobile $\alpha=3/2$  fermions gain extra kinetic energy by maximized expansion (see Fig.~\ref{fig:density_U-8} (b)). 
For sufficiently strong couplings, not only trions but the emerging $m=2$ quintet pairs also become localized in the domains of the quartets.
We refer to this phase as mixed PS-quartetting in Fig.~\ref{fig:phasediagram}, although no sharp transition separates it from the PS quartetting phase. 

In this strong-coupling region, contrary to results for weaker interactions, the weights of the spin components $\alpha=-1/2$, and $-3/2$ vanish simultaneously as the polarization is increased (see Fig.~\ref{fig:densityintegral_U-2} (b)) implying a direct transition between the four- and two-component states.
The two-component system, found for sufficiently large $p$, is gapless and can be characterized by the spatially non-uniform $m=2$ pairing similarly to results for weaker couplings as depicted in Fig.~\ref{fig:phasediagram}.

\begin{figure}[!tb]
  \includegraphics[scale=0.25]{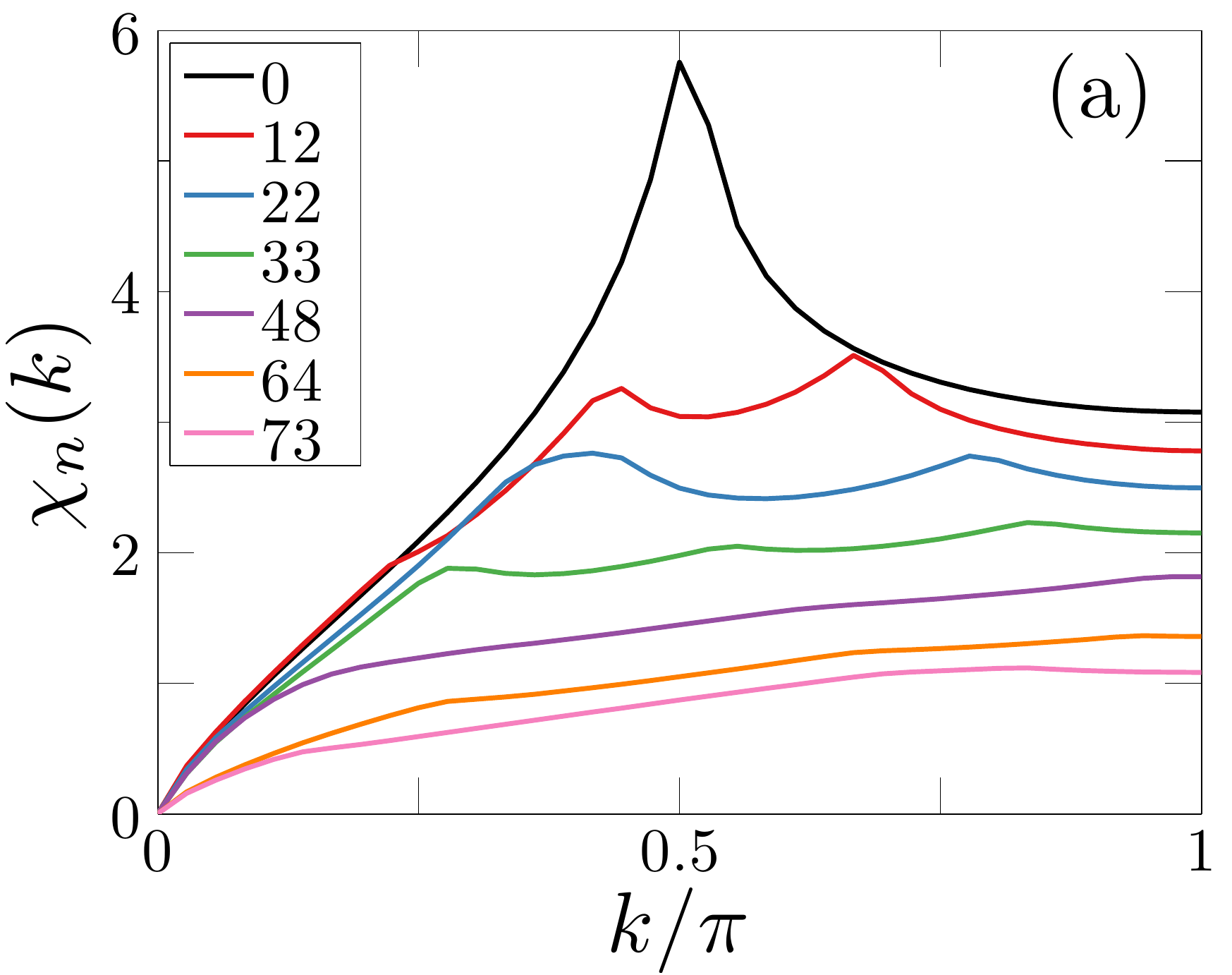}
    \includegraphics[scale=0.25]{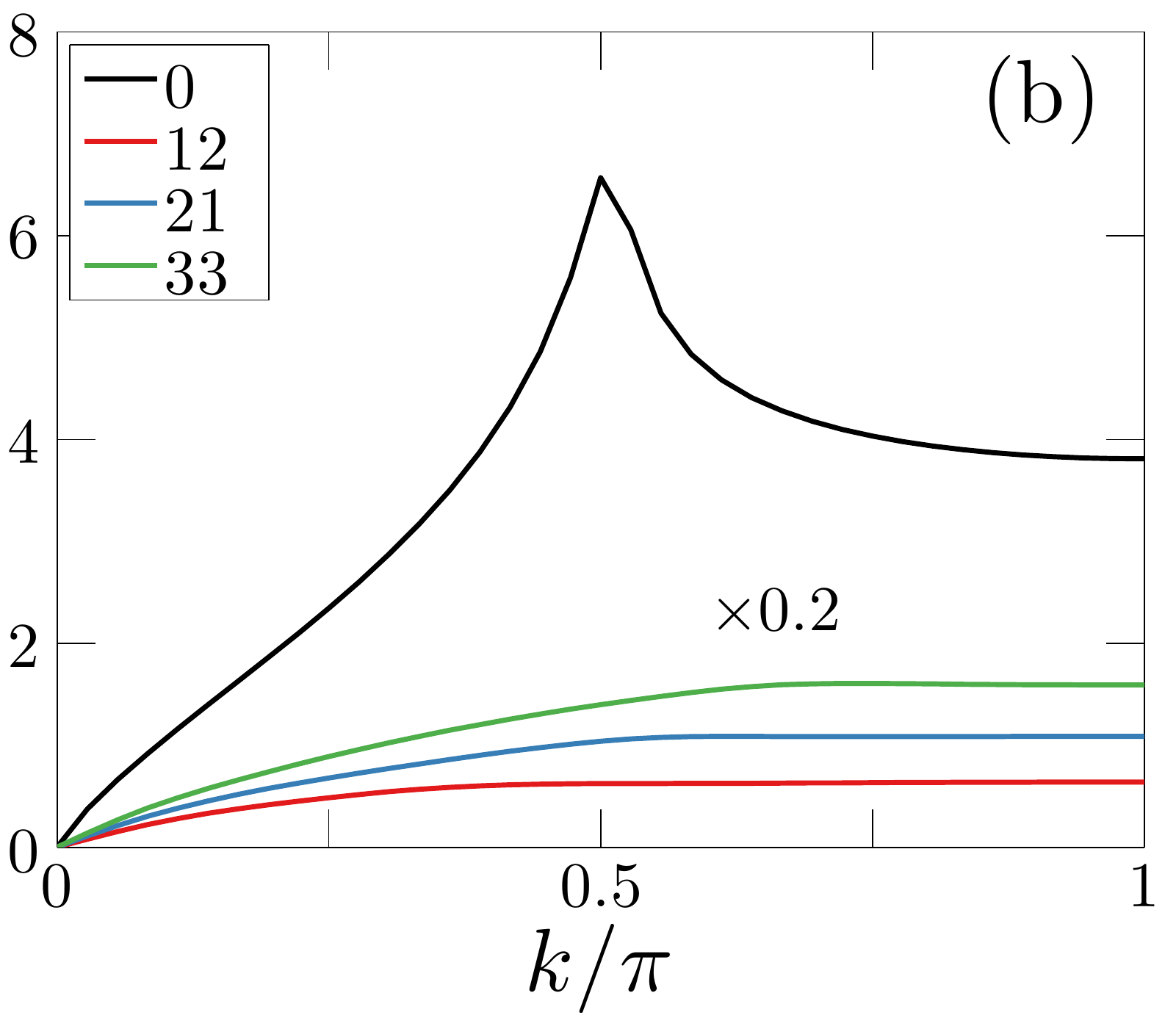}
\caption{(Color online) a) Fourier transform of the density correlations for $U/t=-2$ at various total imbalances $S_z=pL$, denoted with distinct colors. b) Similar for $U/t=-8$,    amplitudes for $S_z \neq 0$ are magnified by a factor of 5 for better visibility. 
}
\label{fig:NNq}
\end{figure}

{\it Observables.} --- The structure of the density can be probed by light-scattering diffraction measurements revealing essentially the Fourier spectrum of the density-density correlations, $\chi_n(k)= 1/L \sum_{l,l'} e^{ik(l-l')} \langle \hat{n}_l \hat{n}_{l'} \rangle$.
In case of spin balance, corroborating previous numerical analysis,\cite{Capponi2007} we find that the oscillation of density correlations develops a quasi-coherent peak at $\pi/2$ (see the $p=0$ curves in Fig.~\ref{fig:NNq} (a) and (b)). 
The characteristic periodicity of the density oscillations, i.e. the position of the quasi-coherent peak, is determined by the Fermi momentum. For finite $p$, the low energy spectrum of the system splits up and the Fermi momenta shift leading to eight different Fermi points as it was discussed in the bosonization section. As a consequence, multiple peaks emerge according to the relative population of the four spin components which can be observed nicely in Fig.~\ref{fig:NNq} (a).

The asymptotic behavior of the density correlations reveals the Luttinger parameter ($K_c$), which is proportional to the compressibility. It can be extracted from the structure factor in the long-wavelength limit, i.e. $K_c=\pi/4 \lim_{k \rightarrow 0} \chi_n(k)/k$. 
We observe that the Luttinger parameter  of the weakly interacting system is independent of the spin imbalance up to critical polarization $p_c \approx 5/6$ where  fermion component $\alpha=-3/2$  freezes out. 
At $p_c$, reaching the  $n=3$ Luttinger liquid phase with strongly asymmetrical filling,  the compressibility of the system drops suddenly to a smaller value  as observed in the $k\rightarrow 0$ limit of the curves in Fig.~\ref{fig:NNq} (a). The value of $p_c$ does not show significant systematic changes for shorter chains either.

The phase separation can also be detected in the changing of $K_c$, i.e. the long-wavelength gradient of the density correlations. We found that $K_c$ is an order of magnitude smaller in the strongly interacting spin-polarized system --- even for small polarization --- compared to the case of spin balance as a direct consequence of the quartet crystallization.
Furthermore, the Luttinger parameter --- and therefore $\kappa$ also --- increases slightly for increasing $p$ as a consequence of the decreasing presence of crystallized quartet domains, and larger weight of the free atoms in the system.

\textit{Concluding remarks.} ---
We investigated a quarter-filled attractive 4-component fermionic system confined into a one-dimensional chain.
While the spin-balanced system is described by quartetting, for finite polarization not only quartets but other particles with different effective mass characterize the ground state.
This mass imbalance can induce the crystallization of heavy molecules. 
In fact, we found that for strong couplings the paramagnetic liquid quartet state collapses into an exotic supersolid-like phase-separated state formed by well localized SU(4) quartets in a background of unbound mobile fermions (PS-quartetting). 
As the polarization is increased, this state becomes richer with further quasi-localized heavy composite molecules. 
It is also observed that for even stronger spin imbalance the $\alpha<0$ fermion components are frozen out simultaneously indicating a direct transition from the mixed PS-quartetting state to the well known pair-FFLO state.
For somewhat weaker interactions an exotic FFLO-like state is observed, where the system is driven from the segregated molecular quartet state to an inhomogeneous mixture of quartets and free $\alpha=3/2$ fermions.

We thank \'A. B\'acsi, G. Tak\'acs, and G. Zar\'and for fruitful discussions.
This research was supported in part by the Hungarian Research Fund (OTKA) under Grant Nos. K~100908, K~105149, and NN110360. 
The authors acknowledge computational support from Philipps Universit{\"a}t, Marburg. E.Sz. also acknowledges support from J\'anos Bolyai Scholarship.

\bibliography{references}

\end{document}